\documentclass[aps,prd,showpacs,preprintnumbers,nofootinbib,floatfix,floats,groupedaddress,twocolumn]{revtex4-1}
\usepackage{bm}
\usepackage{latexsym}
\usepackage{dcolumn}
\usepackage{amsmath,amsfonts,amssymb}
\usepackage{graphicx,epsfig}
\usepackage{color}
\usepackage[active]{srcltx}%DO NOT DELETE
\usepackage{subfig}
\usepackage{slashed}
\usepackage{mathrsfs}
\usepackage{tikz}
\usetikzlibrary{shapes,snakes}
\usepackage{dsfont}
\usepackage[font=normal,labelfont=bf]{caption}
\usepackage{mathtools}
\def\nn{\nonumber}

\def\l{\left}
\def\r{\right}
\def\DM{\mathrm{d}}

\newcommand{\gae}{\lower 3pt \hbox{$\,\, \buildrel {\scriptstyle >}\over {\scriptstyle
\sim}\,\,$}}
\newcommand{\lae}{\lower 2pt \hbox{$\, \buildrel {\scriptstyle <}\over {\scriptstyle
\sim}\,$}}
\reversemarginpar
\def \lp { \ell_0}

\def \ge {\widehat{\bm g}}

\begin{document}

\title{Euclidean Action and the Einstein tensor}

 \author{Dawood Kothawala}
 \email{dawood@iitm.ac.in}
 \affiliation{Department of Physics, Indian Institute of Technology Madras, Chennai 600 036}

\date{\today}
%\date{October 10, 2015}
\begin{abstract}
\noindent
I give a local description of the Euclidean regime $\l( M, \bm g, \bm u \r)$ of Lorentzian spacetimes $\l( M, \bm g \r)$ based on timelike geodesics $\bm u$ passing through an arbitrary event $p_0 \in M$. I show that, to leading order, the Euclidean Einstein-Hilbert action $I_E$ is proportional to the Einstein tensor ${\bm G}\l[\bm g\r]\l(\bm u, \bm u\r)$. The positivity of $I_E$ follows if ${\bm G}\l[\bm g\r]\l(\bm u, \bm u\r)>0$ holds. 
I suggest an interpretation of this result in terms of the amplitude $\mathcal{A}\l[\Sigma_0 \r]=\exp[{-I_E}]$ for a single space-like hypersurface $\Sigma_0 \in I^{+}(p_0)$ to emerge at a constant geodesic distance $\lambda_0$ from $p_0$. Implications for classical and quantum gravity are discussed.
\end{abstract}

\maketitle
\vskip 0.5 in
\noindent
\maketitle
%%%%%%%%%%%%%%%%%%%%%%%%%%%%%%%%%%%%%%%%%%%%%%%%%%%%%%%%%%%%%%%%%%%%%%%%%%%%%%%%%%%%%%%
\section{Introduction} \label{sec:intro} 
%\textbf{\textit{Introduction}}:
%%%%%%%%%%%%%%%%%%%%%%%%%%%%%%%%%%%%%%%%%%%%%%%%%%%%%%%%%%%%%%%%%%%%%%%%%%%%%%%%%%%%%%%
Feynman's path-integral formulation of quantum theory provides a powerful basis for setting up a quantum framework for a theory described by certain degrees of freedom, say $q_A$, with probability amplitudes for different configurations determined by the classical action $I[q_A]$. If one therefore wishes to study the quantum aspects of gravity within the path integral formalism, it is natural to start with the Einstein-Hilbert action on a manifold $\l( M, \bm g \r)$, determined by the lagrangian $L_{\rm grav}[\bm g]=\bm{\mathrm{RicSc}}[\bm g]$ - the Ricci scalar constructed from $\bm g$ (and appropriate boundary term for each boundary of $M$). The corresponding path-integral for gravity is then defined as a sum-over-histories $\bm g$ of the amplitude $\mathcal{A}\l[\mathscr{G}_{\rm f}, \mathscr{G}_{\rm i} | \bm g\r]=\exp\l[ i I\l[\mathscr{G}_{\rm f}, \mathscr{G}_{\rm i} | \bm g\r]/\hbar \r]$ which is the transition amplitude between the $3$-geometries $\mathscr{G}_{\rm i}$ and $\mathscr{G}_{\rm f}$ corresponding to a given $\bm g$ (mathematically, a Lorentzian cobordism). All these steps are merely formal - they simply state the standard prescription of path-integrals for a classical tensor field $\bm g$. But of course, gravity is more than simply a theory of a classical field, it is also a manifestation of the curvature of spacetime which provides the background over which all other field theories are constructed. This makes the situation much more complicated, and has been discussed at length in the vast literature on the topic. 

In this work, I focus on the most basic of these: The Euclidean version of the gravitational path-integral \cite{book-gh}. The conventional approach here is to perform a suitable Wick rotation (analytic continuation of time coordinate $t$ to complex plane), and then study the path-integral based on the lagrangian $L_{\rm grav}[\bm g_E]=\bm{\mathrm{RicSc}}[{\bm g}_E]$, where $\bm g_E$ is the Euclidean metric. Of course, Wick rotation does not always yield a sensible $\bm g_E$, and many variants have been proposed which analytically continue some metric degree of freedom as a cure for issues related to analytic continuation of $t$ and/or those related to the unboundedness of the Euclidean action. Be that as it may, such issues definitely make it worthwhile to probe deeper the class of Euclidean geometries that can be introduced in the path-integral, and that is compatible with the existence of a Lorentzian metric on $M$. 

With this in mind, I here consider a covariant alternative to conventional Wick rotation ($t \to it$), which is essentially motivated by a simple result about existence of Lorentzian metrics on manifolds that possess a Euclidean metric [Sec.\,2.6, Hawking and Ellis \cite{book-he}]. Specifically, a manifold with a Euclidean metric admits a Lorentzian metric  (or the converse, which is more relevant for our case) if there exists a smooth, nowhere vanishing vector field $\bm u$ on it. Such a vector field always exists for non-compact manifolds, while compact manifolds admit one iff their Euler number is zero. I therefore focus on the class of Euclidean metrics
%\begin{eqnarray}
$
%\bm \ge^{-1} = \bm g^{-1} - \Theta \, \bm u \otimes \bm u
{\widehat g}^{ab} = g^{ab} - \Theta(\lambda) \, u^a u^b
$
%\end{eqnarray}
where $u^a$ is a well-defined unit timelike vector field parametrised by $\lambda$ (that is, $g_{ab}u^a u^b=-1$ and $u^a \partial_a \lambda=1$), and $\Theta(\lambda)$ is a transition function that satisfies
$\lim \limits_{x \rightarrow 0} \Theta(x)=-2$
and
$\lim \limits_{x \rightarrow \infty} \Theta(x) = 0$
%\begin{eqnarray}
%\lim \limits_{x \rightarrow 0} \Theta(x) &=& - 2
%\nn \\
%\lim \limits_{x \rightarrow \infty} \Theta(x) &=& \phantom{-} 0
%\end{eqnarray}
corresponding to the metric $\bm \ge$ being Euclidean or Lorentzian -- in particular, $\bm g_E \equiv \ge(\Theta=-2)$ \cite{book-he, candelas-raine-visser-samuel, loll-dg-propertime}.
\begin{figure*}%
    \centering
%    {{\includegraphics[width=0.85\textwidth]{cst4} }}%
    {{\includegraphics[width=0.85\textwidth]{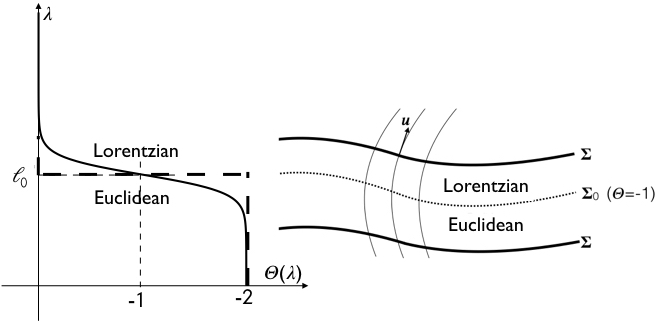} }}%
    \caption{{\it Left}: A typical profile for $\Theta(\lambda)$; the dashed curve is the idealised step profile used in this paper. {\it Right}: Euclidean to Lorentzian transition characterised by $\bm u$ and $\Theta(\lambda)$. $\Sigma$ represent level surfaces of $\bm u$.}%
    \label{fig:theta}%
\end{figure*}
            %\begin{figure*}%
            %    \centering
            %    \subfloat[A typical profile for $\Theta(\lambda)$. The dashed curve is the idealised step profile used in this paper.]{{\includegraphics[width=0.35\textwidth]{fig-thetan} }}%
            %    %%
            %    \subfloat[Euclidean to Lorentzian transition characterised by $\bm u$ and $\Theta(\lambda)$. $\Sigma$ represent level surfaces of $\bm u$, with $\Sigma_0$ representing the one on which $\Theta=-1$.]{{\includegraphics[width=0.6\textwidth]{Euclidean-QG-fig1} }}%
            %    \caption{Euclidean domain of Lorentzian geometries.}%
            %    \label{fig:geodesic-congruence}%
            %\end{figure*}
I will assume that the transition between these two values of $\Theta$ is sharp -- see Fig. (\ref{fig:theta}). Although the two domains -- Euclidean and Lorentzian -- are of primary interest here, the {\it transition} between these also leads to interesting mathematical structure in the curvature tensors, represented by terms with delta function support. Several novel and remarkable consequences follow from this proposal for Euclidean regimes associated with Lorentzian spacetimes \cite{dk-cqgl}, resulting in a rich geometrical structure. {\it As we shall see, combined with the geodesic structure of Riemannian/Lorentzian space(time)s, these features imply a very specific relationship between the Euclidean Einstein-Hilbert action $I_E := - i I[\bm g_E]$ and the Einstein tensor $\bm{\mathrm{G}}[{\bm g}]$}.

Before proceeding to prove this relationship, let me highlight two key advantages of studying Euclidean quantum gravity in the framework proposed here. First, it helps us to define a Euclidean geometry corresponding to a given Lorentzian geometry without any ambiguity and without having to worry about the metric components becoming imaginary. This is in contrast to what happens with conventional Wick rotation. Second, the fact that the resultant {\it Euclideanisation} depends on a vector field $\bm u$ allows us to introduce the notion of {\it observer dependence} at a very basic level in the quantum description of gravity, a desirable feature since quantum theory is expected to be inherently observer dependent (a fact that has not received as much careful attention as other aspects of quantum gravity, though some discussions exist; for e.g., see \cite{calzetta-kandus}).

%%%%%%%%%%%%%%%%%%%%%%%%%%%%%%%%%%%%%%%%%%%%%%%%%%%%%%%%%%%%%%%%%%%%%%%%%%%%%%%%%%%%%
\section{The curvature tensors associated with $\ge$} \label{sec:curvature} 
%\textbf{\textit{The curvature tensors associated with $\ge$}}:
%%%%%%%%%%%%%%%%%%%%%%%%%%%%%%%%%%%%%%%%%%%%%%%%%%%%%%%%%%%%%%%%%%%%%%%%%%%%%%%%%%%%%
I will now describe the geometrical features associated with the metric $\bm \ge$ that will allow us to construct the action $I[\bm \ge]$, whose Euclidean regime will be our key point of focus.
After lengthy algebra and judicious use of Gauss-Codazzi and Gauss-Weingarten equations, it is possible to write down the geometrical quantities associated with $\bm \ge$ in terms of those associated with $\bm g$. This inevitably involves the intrinsic and extrinsic geometry of $\bm u$ foliation with the induced metric (the projection of) $h^a_{\phantom{a} b}=\delta^a_{\phantom{a} b} + u^a t_{b}$. (Here, $t_a=g_{ab}u^b$.) 
Some relevant expressions are given in the Appendix for completeness; these lead to the final expression for the Ricci scalar which is of direct relevance for further discussion of the Euclidean action
\begin{eqnarray}
%$$
\bm{\mathrm{RicSc}}[\widehat{\bm g}] =  \l(1 + \Theta \r) \bm{\mathrm{RicSc}}[\bm g] - \Theta \; {\mathcal R}_{\Sigma} +  \l( \frac{\DM \Theta}{\DM \lambda} \r) K 
\hspace{1cm}
\label{eq:ricci-scalar-gen}
%$$
\end{eqnarray}
where ${\mathcal R}_\Sigma$ represents the intrinsic Ricci scalar of level surfaces of $\bm u$ (see Fig. (\ref{fig:theta})), and $K$ their extrinsic curvature.
We will now use the above to evaluate the action in the Euclidean regime of $\widehat{\bm g}$. For this, we will choose a sharp (step-function) profile for the transition function $\Theta(\lambda) = 2 \theta(\lambda-\lambda_0) - 2$. 
%(We are essentially ignoring the {\it width} of transition, which we do not expect to cause any significant conceptual change.) 
Since $\DM \Theta/\DM \lambda = 2 \delta(\lambda-\lambda_0) \equiv 2 \delta_{\Sigma_0}$, the last term in the above expression will contribute 
$(2 K) \delta_{\Sigma_0}$ to the Euclidean action, which happens to be precisely the Gibbons-Hawking-York (GHY) boundary term in $D=4$! This somewhat curious result arises because the metric signature changes by $2$ (which leads to the correct factor of $2$ in the GHY term).

We are now in a position to analyse the action 
\begin{eqnarray}
\frac{I[\widehat{\bm g}]}{\hbar} = \frac{1}{\lp^{D-2}} \int \bm{\mathrm{RicSc}}[\widehat{\bm g}] \; \DM v_{D}
\label{eq:action}
\end{eqnarray}
where $\DM v_{D}$ is the volume measure based on $\widehat{\bm g}$ 
\footnote{$\lp$, with dimensions of length, is {\it defined} by this expression (it is the natural relativistic reduced Planck scale). To avoid clutter, we will set $\lambda_0=\lp$, since we expect the transition to take place close to Planck scale. It is easy to do away with this choice, in which case the ratio $(\lambda_0/\lp)$ will appear in the final result.}
. We will be interested in the Euclidean regime $\Theta=-2$, and hence the volume integration will be over the corresponding domain. Finally, we note that, ${\rm det}\, \bm \ge = (1+\Theta)^{-1} \, {\rm det}\, \bm g$, and since ${\rm det}\, \bm g<0$, $\sqrt{- {\rm det}\, \bm \ge}$ is imaginary for $\Theta < -1$, and in particular for $\Theta=-2$. This is expected. However, for $\Theta=-1$, the metric $\widehat{\bm g}$ is degenerate (in fact, equal to $h_{ab}$). Therefore, we shall choose the volume measure $\DM v_{D}$ as equal to $\sqrt{- {\rm det}\, \bm \ge} \; \DM^4x=i \sqrt{- {\rm det}\, \bm g}\; \DM^4x$ for 
$\Theta<-1$, and equal to $\sqrt{-{\rm det}\, \bm h} \; \DM^4x=i \sqrt{{\rm det}\, \bm h} \; \DM^4x$ for $\Theta=-1$.

\begin{figure*}%
    \centering
%    {{\includegraphics[width=0.85\textwidth]{cst3} }}%
    {{\includegraphics[width=0.85\textwidth]{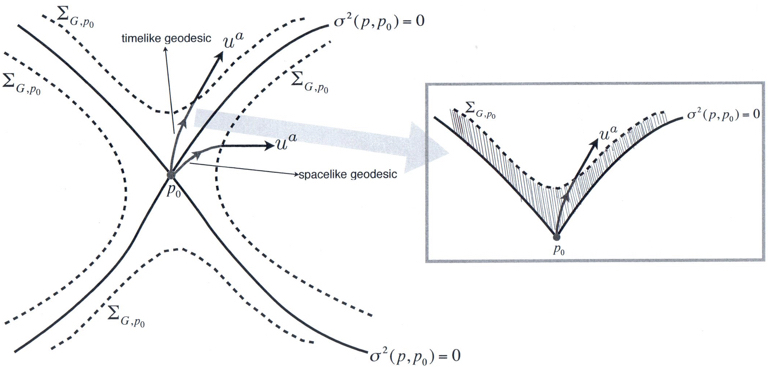} }}%
    \caption{Geodesic structure of spacetime near an arbitrary event $p_0$. {\it Right inset}: Future timelike geodesics in $I^+(p_0)$ will serve as the basis for the local Euclidean regime in the neighbourhood of $p_0$. The shaded region represents the Euclidean domain.}%
    \label{fig:cst}%
\end{figure*}
%%%%%%%%%%%%%%%%%%%%%%%%%%%%%%%%%%%%%%%%%%%%%%%%%%%%%%%%%%%%%%%%%%%%%%%%%%%%%%%%%%%%%
\section{The local Euclidean geometry of spacetime} \label{sec:euc-geom} 
%\textbf{\textit{The local Euclidean geometry of spacetime}}:
%%%%%%%%%%%%%%%%%%%%%%%%%%%%%%%%%%%%%%%%%%%%%%%%%%%%%%%%%%%%%%%%%%%%%%%%%%%%%%%%%%%%%
We are now ready to study the Euclidean regime in a geodesically convex neighbourhood of an arbitrary event $p_0$ in a manifold possessing a Lorentzian metric $\bm g$, using for $\bm u$ the set of timelike geodesics emanating from $p_0$. Our construction, being anchored at an (otherwise arbitrary) spacetime event $p_0$ and valid within $I^+(p_0)$, therefore provides a {\it local}, {\it covariant} prescription for Euclidean action as an alternative to conventional Wick rotation.
 
Since $\bm u$ are timelike geodesics emanating from $p_0$, the surfaces of constant geodesic distance along $\bm u$ are orthogonal to $\bm u$; see Lemma $4.5.2$ of Hawking and Ellis \cite{book-he}. The corresponding surfaces, which we call {\it equi-geodesic} surfaces, then represent $\Sigma$, and comprise of events $p$ lying at constant (squared) geodesic interval $\sigma^2(p,p_0)$ from $p_0$. The relevant geometrical properties of such surfaces in arbitrary curved spacetimes were discussed in \cite{equi-geod}, and we briefly quote the results which we will need here. First, it is easy to show that $t_a = {\nabla_a \sigma^2}/{2 \sqrt{- \sigma^2}}$. From this, the extrinsic curvature of $\Sigma$ can be computed as $K_{ab} = \l(- \sigma^2\r)^{-1/2} \l( \nabla_a \nabla_b \l( \sigma^2/2 \r) + t_a t_b \r)$. All the interesting geometric properties of $\Sigma$ can therefore be derived from the well known covariant Taylor series expansion of the bi-tensor 
$\nabla_a \nabla_b \l( \sigma^2/2 \r)$ at $p$ near $p_0$ \cite{christenson-worldf}. The quantities of relevance to us have the following covariant Taylor expansions (in $\lambda=\sqrt{- \sigma^2}$) characterised essentially by the {\it tidal tensor} $\mathcal E_{ab} = R_{a m b n} u^m u^n$
\begin{eqnarray}
K &=& {D_1}/\lambda - (1/3) \lambda \mathcal{E} + (1/12) \lambda^2 \nabla_{\bm u} \mathcal{E} - (1/60) \lambda^3 \mathcal{F} + O(\lambda^4)
\nn \\
\mathcal R_{\Sigma} &=& -{D_1 D_2} \; \lambda^{-2} + R + (2/3)(D+1) \mathcal{E} + O(\lambda)
\label{eq:taylor-exp}
\end{eqnarray}
where $\mathcal{E}=g^{ab}\mathcal E_{ab}$, $\mathcal{F} = \nabla_{\bm u}^2 \mathcal{E} + (4/3) \mathcal{E}^a_b \mathcal{E}^b_a$, and we use the convenient shorthand $D_{\#}$ to denote $D-{\#}$.

%%%%%%%%%%%%%%%%%%%%%%%%%%%%%%%%%%%%%%%%%%%%%%%%%%%%%%%%%%%%%%%%%%%%%%%%%%%%%%%%%%%%%
\section{The Euclidean action} \label{sec:euc-action}
%\textbf{\textit{The Euclidean action}}:
%%%%%%%%%%%%%%%%%%%%%%%%%%%%%%%%%%%%%%%%%%%%%%%%%%%%%%%%%%%%%%%%%%%%%%%%%%%%%%%%%%%%%
To evaluate the Euclidean action, it is convenient to write the Lorentzian metric $\bm g$ at events $p \in I^+(p_0)$ in the {\it synchronous} coordinates: $\bm g= - \bm \DM \lambda \otimes \bm \DM \lambda + \bm h(\lambda, \chi, \Omega^A)$, where $\chi$ is the local boost coordinate and $\Omega^A, A=3 \ldots D$ are angular coordinates. It is easy to show that ${\rm det} \bm h$ has the following expansion in $\lambda$: $\sqrt{{\rm det}\, \bm h}\; \DM \chi \DM \Omega^A = \lambda^{D-1} \left[ 1 - (1/6) \mathcal{E} \lambda^2 + O(\lambda^3)\right] (\sinh{\chi})^{D-2} \; \DM \chi \DM \Omega^A$. 
\footnote{Note that $\mathcal E=\mathcal E_{ab}(p_0) u^a(\chi, \Omega^A) u^b(\chi, \Omega^A)$, though we will suppress the dependence on $(\chi, \Omega^A)$ to avoid notational clutter.}

We can now use Eq.~(\ref{eq:ricci-scalar-gen}) with $\Theta=-2$, along with (\ref{eq:taylor-exp}), the expression for $\DM v_D$ (discussed below Eq.~(\ref{eq:action})), and the above expansion for $\sqrt{{\rm det}\, \bm h}$, to evaluate the Euclidean action. The $\lambda$ integral goes from $\lambda=0$ to $\lambda=\lp$, and keeping in mind the $(2 K) \delta_{\Sigma_0}$ term, a lengthy computation finally yields (recall that $I_E := -iI[\bm g_E]$)
\begin{eqnarray}
\frac{I_E}{\hbar} = \frac{1}{D} \int \lp^2 
\l[ R + \frac{1}{3} \l( D_1 D_2 - D_{-1}D_4 \r) {\mathcal E} \r]
\DM \mathds{H}^{D-1}_1 
\nn \\
+ \underbrace{ O(\lp^3 \times {\nabla \mathcal R} \ldots)}_{\rm higher~curvature~terms}
\nn
\end{eqnarray}
which, upon using $\mathcal E = R_{ab} u^a u^b = G_{ab} u^a u^b - (1/2)R$, simplifies remarkably, thereby yielding our key result
\begin{eqnarray}
\frac{I_E}{\hbar} &=& \frac{2}{D} \int \lp^2 G_{ab}(p_0) u^a u^b \DM \mathds{H}^{D-1}_1 \; + \underbrace{ O(\lp^3 \times {\nabla \mathcal R} \ldots)}_{\rm higher~curvature~terms}
\nn \\
&\approx& \frac{2}{D} \lp^2 G_{ab}(p_0)\, \tau^{ab} %\mathds{T}^{ab}
\label{eq:actionE}
\end{eqnarray}
where $\tau^{ab}=\int u^a u^b \DM \mathds{H}^{D-1}_1$ represent the average of unit timelike vectors $u^a(\chi, \Omega^A)$ over the unit $(D-1)$ hyperbolic space $\mathds{H}^{D-1}_1$. In particular, it is evident that, as long as ${\bm G}\l[\bm g\r]\l(\bm u, \bm u\r)>0$ for all timelike vectors $\bm u$, $I[\bm g_E] >0$.

This is a remarkable result, and the only inputs that have gone into deriving this result are (i) the characterisation of $\ge$, and (ii) geometry of level surfaces of timelike geodesics emanating from a spacetime event $p_0$. Both of these inputs are rooted in basic differential geometry (see, for e.g., \cite{book-he}), and provide a more rigorous alternative to Wick rotation for studying Euclidean regime of spacetime. Irrespective of how one proceeds further from it, Eq.~(\ref{eq:actionE}), which is our main result, is sufficient to indicate the non-trivial role that the Einstein tensor of a given Lorentzian geometry plays in determining the structure of the action in the Euclidean regime of this geometry. To the best of my knowledge, such a connection has neither been expected nor arrived at in the conventional approach to Euclidean quantum gravity.

Formally, of course, $\tau^{ab}$ is divergent due to the exponentially divergent volume of the hyperbolic space. I briefly mention below two possible ways for evaluating $\tau^{ab}$. Although both are mathematically straightforward, I must add that there is no preferred way of choosing one over other without entering into the realm of speculations. It is not even clear whether one should bother about it at this stage, since it is the Euclidean path integral based on $I_E$ which is expected to be more relevant than $I_E$ itself. 
%(On a somewhat different note, which one is {\it termed} as mathematically more rigorous depends, unfortunately, on considerations other than {\it mathematical rigor}!):

(a) {\it Imposing cut-off on $\chi$}: The most straightforward evaluation is done by replacing $\int_0^\infty \DM \chi (\ldots) \to \int_0^{\chi_c} \DM \chi (\ldots)$ to extract the leading $\chi_c \to \infty$ divergences. The evaluation for $\tau^{ab}$ in this case is most conveniently done by parametrising $u^a$ with standard Lorentz transformations: $u^a(\chi, \Omega^A) = \l( \cosh{\chi} \r) T^a + \l( \sinh{\chi} \r) N^a$, where $T^a, N^a$ are arbitrary unit timelike, spacelike vectors in the tangent space $\mathcal{T}_{p_0}(M)$, with $T^a N_a=0$. It is then straightforward to show that $\tau^{ab}/S_{D-2} = (I_D/(D-1)) \l[ \eta^{ab} + D T^a T^b \r] + I_{D-2} T^a T^b$, where $S_{D-2}$ is the volume of unit $(D-2)$ sphere. In this form, the $\chi_c \to \infty$ divergences are captured through the integrals $I_D = \int_0^{\chi_c} \DM \chi \, (\sinh \chi)^D$. It is worth highlighting that the first term in the structure of $\tau^{ab}$, being traceless, would pick the traceless part $G^{\rm tr}_{ab}=G_{ab}-(1/D)Gg_{ab}$ of the Einstein tensor. Explicitly, $G_{ab} \tau^{ab}/S_{D-2} = (D I_D/(D-1)) G^{\rm tr}_{ab} T^a T^b + I_{D-2} G_{ab} T^a T^b$. The Euclidean action with this regularisation is worth exploring further, and can lead to new insights into quantum gravity as well as its classical limit. It might also be of direct conceptual significance for ideas that treat gravity as an emergent phenomenon \cite{eg-jpv}.

(b) {\it Regularised hyperbolic volume}: As an alternative to the above regularisation, one might mention that there has been discussion on handling precisely the above kind of divergences in the context of AdS-CFT, which essentially regularises the volume of $\mathds{H}^N$ (which is exactly what arises in our setup as well). It is straightforward to show that $\tau^{ab} \equiv ({\rm vol_{reg}}(\mathds{H}^{D-1}_1)/D) g^{ab}(p_0)$ \cite{maldacena-reg-vol}. In this case, 
$G_{ab} \tau^{ab} = ({\rm vol_{reg}}(\mathds{H}^{D-1}_1)/D) G = - {\rm vol_{reg}}(\mathds{H}^{D-1}_1) ((D-2)/2D) \bm{\mathrm{RicSc}}[\bm g]$. The Euclidean action is now indeed proportional to $\bm{\mathrm{RicSc}}[\bm g]$, but the 
proportionality constant is not the standard one. The relevance and/or justification for this particular regularisation is unclear (at least to this author).

%%%%%%%%%%%%%%%%%%%%%%%%%%%%%%%%%%%%%%%%%%%%%%%%%%%%%%%%%%%%%%%%%%%%%%%%%%%%%%%%%%%%%
\section{Discussion and Implications} \label{sec:implications} 
%\textbf{\textit{Discussion and Implications}}:
%%%%%%%%%%%%%%%%%%%%%%%%%%%%%%%%%%%%%%%%%%%%%%%%%%%%%%%%%%%%%%%%%%%%%%%%%%%%%%%%%%%%%
Let me first summarise the approach presented here and the result it has led us to. I began by considering a class of spacetime metrics $\ge$ derivable from a Lorentzian metric $\bm g$ and timelike geodesics $\bm u$, which interpolate between the Euclidean and Lorentzian space(time)s. This turns out to lead to a rich mathematical structure, with the transition between Euclidean and Lorentzian regimes leading to terms in curvature with delta function support on the hypersurface on which the transition takes place. Even more surprisingly, the Ricci scalar $\bm{\mathrm{RicSc}}[\widehat{\bm g}]$ corresponding to $\ge$ has a delta function term which corresponds precisely to the GHY boundary term in the conventional formalism of the Einstein-Hilbert action principle. In addition, $\bm{\mathrm{RicSc}}[\widehat{\bm g}]$ in the Euclidean regime has an additional term involving intrinsic Ricci scalar of the co-dimension one transition surface. This entire formalism is then applied to the causal future of an arbitrary spacetime event $p_0$, using for $\bm u$ the timelike geodesics emanating from $p_0$. This yields a local description of Euclidean regime in the neighbourhood of any event $p_0$. I then computed the Euclidean action $I_E$ explicitly and exhibited it's direct connection with the Einstein tensor of $\bm g$.

{\it I must emphasize that the connection between the {\it Euclidean} action and the {\it Lorentzian} Einstein tensor, derived here, is a highly non-trivial result and there seems to be no a priori reason for expecting such a connection.} 
\footnote{One plausible connection is hinted by the case of static solutions in standard field theories. Here, it is well known that the Euclidean action is the Hamiltonian (apart from a factor of the periodicity of euclidean time). Since $G^0_0$ is essentially the gravitational Hamiltonian, the connection with Euclidean action seems plausible. However, for static solutions in general relativity, the situation can be more subtle \cite{hawking-horowitz}. Moreover, the result derived here does not assume staticity etc. I nevertheless thank the referee for bringing this interesting point to my notice.}
Since it uses covariant expansions valid in arbitrary Lorentzian spacetimes, the result has direct implications for studying quantum properties of the small scale structure of spacetime (perhaps along the lines of \cite{equi-geod, dk-minimal-length}). Let me elaborate a little bit on this, taking cue from the domain in which similar ideas from Euclidean quantum gravity were first applied and developed - {\it quantum cosmology}. We will focus on the well known Hawking-Hartle prescription for the ground state wave function of the universe \cite{book-gh, hhWfn}. This is defined via the path integral over Euclidean geometries that have a $(D-1)$ hypersurface $\Sigma_0$ as their only boundary, and, in the semi-classical limit, the corresponding wave function  $\bm \Psi[\Sigma_0]$ is interpreted as yielding the amplitude for the universe to emerge from nothing. With this as motivation, we may consider the result here derived as yielding a wave function $\bm \Psi_{p_0} \sim e^{-(2/D) \lp^2 G_{ab} \tau^{ab}}$ describing emergence of a single space like surface at a fixed geodesic distance from an arbitrary event $p_0$. Since the analysis is completely local, one may then apply it all of spacetime, in which case one would then be effectively talking about the wave function $\Psi = \underset{p_0}{\Pi} \Psi_{p_0}$ for a spacetime with a given Lorentzian metric $\bm g$ to exist. Understanding of our result along these lines would also then pave way to understand better the role of an {\it observer} as far as the small scale structure of spacetime is concerned, somewhat along the lines of Calzetta and Kandus \cite{calzetta-kandus}, who argued that quantum cosmology inherits the observer dependence of vacuum in quantum field theory (in their case, through the choice of Wick rotation).

%%%%%%%%%%%%%%%%%%%%%%%%%%%%%%%%%%%%%%%%%%
% Figure explaining speculation motivated by Hartle-Hawking no-boundary proposal
%%%%%%%%%%%%%%%%%%%%%%%%%%%%%%%%%%%%%%%%%%
%        \begin{figure*}
%            \centering
%            {{\includegraphics[width=.85\textwidth]{FIG-hh} }}
%            \caption{Wave function of spacetime? A proposed interpretation of the result derived in this Letter along the lines of Hartle-Hawking prescription for the wave function of the Universe.}
%            \label{fig:hh}
%        \end{figure*}
%%%%%%%%%%%%%%%%%%%%%%%%%%%%%%%%%%%%%%%%%%
%%%%%%%%%%%%%%%%%%%%%%%%%%%%%%%%%%%%%%%%%%

It would also be of interest to understand implications of the result derived here for the {\it positive action conjecture} in Euclidean gravity, and its connection with the energy conditions of classical general relativity. Such a connection is hinted by the proportionality derived here (to leading order in curvature) between $I_E$ and $\bm G[\bm g](\bm u, \bm u)$, since $\bm G[\bm g](\bm u, \bm u) \geq 0$ is (the geometrical version of) the {\it weak energy condition}. (There is already a connection between {\it positive energy theorem} in $(D+1)$ dimensions and {\it positive action conjecture} in $D$ dimensions - the former implies the latter \cite{positive-action-energy}.) 

Finally, as is evident, the result presented here has obvious relevance to quantum gravity, particular those frameworks that use the gravitational path integral as their basic starting point (e.g., Causal Sets, Causal Dynamical Triangulation (CDT)). One would like to study the partition function $Z$ for quantum gravity based on the class of space(time)s described by $\bm \ge$:
$$
%\begin{eqnarray}
Z = \int \mathcal{D} \bm g \; \mathcal{D} \bm u \; \exp{\l[ + i\int \widehat{R} \sqrt{- {\rm det} \bm \ge} \r]}
%\end{eqnarray}
$$
Studying the behaviour of this path integral in the Euclidean regime of $\bm \ge$ should yield new insights, since the integrand in that limit directly depends on the Einstein tensor $G_{ab}$. This would entail addressing several issues, conceptual as well as mathematical, so as to fully extract the consequences of the result for small scale structure of spacetime. It would also be of interest to investigate the effective action obtained by integrating over $\bm u$. Although the full treatment of this might be involved, in the limit being considered, since the euclidean action becomes quadratic in $\bm u$, the path integral can presumably be done (and will be determined by the determinant of the Einstein tensor $\bm G$). However, it is best not to speculate about this without further careful consideration of the higher curvature terms in the (euclidean) action $I_{\rm E}$. 

Of particular interest is the question as to whether the semiclassical limit of $Z$ has any connection with the so called {\it entropy functional formalism} of gravitational dynamics \cite{entropy-functional-tpap}, and, more broadly, for any of the results for the so called {\it emergent gravity paradigm} \cite{eg-jpv}. Since the path amplitudes (to leading order in curvature expansion) are given by $\bm \Psi_{p_0} \sim e^{-(2/D) \lp^2 G_{ab} \tau^{ab}}$, the Lorentzian metrics $\bm g$ satisfying $\bm G[\bm g]=0$ - {\it the vacuum Einstein equations} - would dominate the path-integral. It would be interesting to make this connection mathematically rigorous after including the matter coupling. Such a possibility is also very strongly suggested by the fact that, for a canonical matter action quadratic in first derivatives, $\mathcal L_{\rm matter}[\bm g_E] =  - T_{ab} u^a u^b$! Also of interest in this context is the understanding of the {\it cosmological constant} \cite{eg-cc} as a low energy relic of small scale structure of spacetime. Moving on to quantum gravity, a natural next step would be to see if any of the existing frameworks lead naturally to $\bm \ge$ (perhaps as an effective metric). Indeed, the notion of signature change at small scales has appeared in several quantum gravity frameworks (see \cite{qg-refs-1, qg-refs-2} for examples from Loop Quantum Cosmology and CDT). The result derived here, being applicable for arbitrary curved spacetimes $(M, \bm g)$, should therefore provide a useful tool for a mathematically rigorous discussion of such a change in quantum spacetime. %These issues are under investigation.
\\
%%%%%%%%%%%%%%%%%%%%%%%%%%%%%%%%%%%%%%%%%%%%%%%%%%%%%%%%%%%%%%%%%%%%%%%%%%%%%%%%%%%%%%%
{\it Acknowledgements}: I would like to thank T. Padmanabhan for comments on the work. The support of Department of Science and Technology (DST), India, through its INSPIRE Faculty Award, is also gratefully acknowledged.
%%%%%%%%%%%%%%%%%%%%%%%%%%%%%%%%%%%%%%%%%%%%%%%%%%%%%%%%%%%%%%%%%%%%%%%%%%%%%%%%%%%%%

\textbf{\textit{Appendix}}:
%%%%%%%%%%%%%%%%%%%%%%%%%%%%%%%%%%%%%%%%%%%%%%%%%%%%%%%%%%%%%%%%%%%%%%%%%%%%%%%%%%%%%
I quote here the expression for Riemann tensor for $\bm \ge$:
$$
\widehat{R}^{ab}_{cd} = {R}^{ab}_{cd} + 2 \Theta \Biggl[ t_m R^{m[a}_{\phantom{m[a}cd} u^{b]} + K^{[a}_{\phantom{[a}[c} K^{b]}_{\phantom{b]}d]} \Biggl] + 2 \dot \Theta u^{[a} K^{b]}_{\phantom{b]}[c} \; t_{d]}
$$
($\dot \Theta=\DM \Theta/\DM \lambda$) from which all other tensors, including the Ricci scalar quoted in the text, can be obtained in a straightforward manner \cite{dk-cqgl}. It is also worth highlighting the following limit on the hypersurface $\Theta=-1$, where the metric (expectedly) becomes degenerate: $\lim \limits_{\Theta \rightarrow -1} \widehat{R}^{ab}_{\phantom{ab}cd} \, e^{(\mu)}_a e^{(\nu)}_b e^c_{(\rho)} e^d_{(\sigma)} = {\mathcal R_{\Sigma}}^{\mu \nu}_{\phantom{\mu \nu}\rho \sigma}$ yielding similar limits for all the other tensors; for e.g., $\lim \limits_{\Theta \rightarrow -1} \widehat{R} = {\mathcal R}_\Sigma$.
%%%%%%%%%%%%%%%%%%%%%%%%%%%%%%%%%%%%%
\widetext
%%%%%%%%%%%%%%%%%%%%%%%%%%%%%%%%%%%%%

%%%%%%%%%%%%%%%%%%%%%%%%%%%%%%%%%%%%%
\end{document}